\documentclass[letter]{aa}  
\usepackage[dvips]{graphicx}
\usepackage{txfonts}
\usepackage{natbib}
\bibpunct{(}{)}{;}{a}{}{,} % to follow the A&A style
\begin{document}

   \title{Spicule emission profiles observed in \ion{He}{i} 10830 \AA}

\author{B.~S\'anchez-Andrade~Nu\~no \inst{1,2}
 \and
   R.~Centeno \inst{3,5}
 \and
    K.~G.~Puschmann \inst{2}
 \and
   J.~Trujillo~Bueno \inst{2,3,4}
 \and
  \\ J.~Blanco~Rodr\'\i guez \inst{2}
 \and
   F.~Kneer \inst{2}
}
\offprints{B.~S\'anchez-Andrade~Nu\~no, 
  \email{bruno@astro.physik.uni-goettingen.de}}

 \institute{Max-Planck-Institut f\"ur Sonnensystemforschung, Max-Planck-Str. 2, D-37191 Katlenburg-Lindau,  Germany
  \and
  Institut f\"ur Astrophysik, Friedrich-Hund-Platz 1, D-37077 G\"ottingen, Germany \\
  \email{bruno@astro.physik.uni-goettingen.de}
  \and
  Instituto de Astrof\'\i sica de Canarias, C/ V\'\i a L\'actea, s/n, E-38205 La Laguna, Tenerife, Spain
  \and
  Consejo Superior de Investigaciones Cient\'ificas, Spain
  \and
  High Altitude Observatory (NCAR), 3080 Center Green Dr. (CG-1), Boulder 80301 CO, USA
}

   \date{Received date; accepted date}

% \abstract{}{}{}{}{} 
% 5 {} token are mandatory
 \abstract
   {}
 {Off-the-limb observations with high spatial and spectral resolution will
help us understand the physical properties of spicules in the solar chromosphere.} 
 {Spectropolarimetric observations of spicules in the \ion{He}{i} 10830\,\AA\ multiplet
     were obtained with the Tenerife Infrared Polarimeter on the
     German Vacuum Tower Telescope at the Observatorio del Teide (Tenerife, Spain). The analysis shows the variation of the  off-limb  
emission profiles as a function of the distance to the visible solar limb. The 
ratio between the intensities of the blue and the red components of this 
triplet $({\cal R}=I_{\rm blue}/I_{\rm red})$ is an observational signature 
of the optical thickness along the light path, which is 
related to the intensity of the coronal irradiation. }
 {We present observations of the intensity profiles of spicules above a
     quiet Sun region. The observable ${\cal R}$ as a function of the
     distance to the visible limb is also given. We have compared our observational
     results to the intensity ratio obtained  
from detailed radiative transfer calculations in semi-empirical models 
of the solar atmosphere assuming spherical geometry. The agreement is purely
     qualitative. We argue that future models of the solar chromosphere and  
transition region should account for the observational constraints 
presented here.
}
   {}


   \maketitle
%
%________________________________________________________________

\section{Introduction}

The {  solar} chromosphere is in a highly dynamic state. Its small-scale structures
evolve in timescales of minutes or even less. 
Due to its low density, the chromosphere is transparent in most of the optical
spectrum. Nevertheless, in lines such as H$\alpha$ and He {\sc i} 10830\,\AA, the
significant absorption provides a mean for direct studies of the chromosphere's
peculiar characteristics, such as bright plages in active regions, dark and
bright fibrils on the disc in both the quiet Sun's network and around
sunspots, as well as spicules above the limb. Recent studies
\citep[e.g.][]{tziotziou03} claimed that it is  possible that many of these chromospheric
features have the same physical nature within different scenarios.  

In the chromosphere, between the photosphere and the much hotter
corona, the average temperature starts to rise outwards, along with a
decrease of the density. The dynamics of a magnetised gas depends on  the ratio of the gas pressure and magnetic pressure, the plasma
$\beta$ parameter,
$\beta$\,=\,$P_\mathrm{gas}/P_\mathrm{mag}$, with
$P_\mathrm{mag}$\,=\,$B^2/(8\pi)$. From the low chromosphere into the
extended corona, this plasma parameter decreases from values $\beta>1$ to a
low-beta regime, $\beta\ll1$, where the plasma motions are magnetically driven.

Spicules, known for more than 130 years \citep[see hand drawings in][]{secchi1877}, represent a prominent example of the dynamic
chromosphere. We refer the reader to reviews by \citet{beckers68,beckers72} and to the paper by
\citet{wilhelm00} on UV properties. Spicules are seen at and outside the limb
of the Sun as thin, 
elongated features that develop speeds of 10--30 km\,s$^{-1}$ and reach
heights of 5--9 Mm on average, during their lifetimes of 3--15 minutes. As pointed out by \citet{sterling00}, a key impediment to develop a satisfactory understanding has been the 
lack of reliable observational data. 
Many theoretical models have been developed to understand the 
nature of spicules, using a wide variety 
of motion triggers and driving mechanisms. 
In this study we focus on the \ion{He}{i} 10830\,\AA\ triplet emission line using recent 
technical improvements in observational facilities. We are able to provide 
observational evidence of the link between the corona and the infrared 
emission of this line, in the frame of the current theoretical models of the
solar atmosphere.

The \ion{He}{i} 10830~\AA\ multiplet consists of the three transitions from the upper term {  $^3$P$_{2,1,0}$}, which has three levels,  to the lower {  metastable} term {  $^3$S$_{1}$}, which has one single level. The two transitions from the J=2 and J=1 upper levels appear blended at typical chromospheric temperatures, and form the so-called red component, at
10830.3 \AA. {  (Note that the two red
 transitions are only 0.09~\AA\ apart).}  The blue component, at 10829.1 \AA,  corresponds to the
transition from the upper level with J=0 {  to the lower level with J=1}. 
The energy levels that take part in these transitions are basically populated via
an ionization-recombination process \citep{avrett94}. The EUV coronal irradiation (CI) at 
wavelengths lambda $\lambda<504$~\AA\ ionizes the neutral helium, and subsequent recombinations
of singly ionized helium with free electrons lead to an overpopulation of these
levels.

{  Alternative theories suggest other mechanisms that might also contribute to the formation of the helium lines relying on the collisional excitation of the electrons in regions with higher temperature \citep[e.g.,][]{a97}}

\citet{Centeno06} modelled the ionisation and recombination processes using various   
 amounts of CI, non-LTE radiative transfer, and different atmospheric
 models {  \citep[see also][in preparation]{cente07}}.
 They have simulated limb observations for different heights, obtaining
 synthetic emission profiles in spherically symmetric models of the solar atmosphere. One
 conclusion of their 
 study is that the ratio of intensities $({\cal R}=I_{\rm blue}/I_{\rm red})$
 of the `blue' to the `red' components of the \ion{He}{i} 10830~\AA\ emission
 is a very good candidate for diagnosing the CI.
 The population of the metaestable level depends on optical thickness, 
whose variation with height governs the change in the ratio $\cal R$ as 
a function of the distance to the limb.
 
\citet{truj05} measured the four Stokes parameters of quiet-Sun chromospheric spicules and could show evidence of the Hanle effect by the action of inclined magnetic fields with an average strength of the order of 10 G. They modelled the \ion{He}{i} 10830~\AA\
profiles assuming the medium along the integrated line of sight as
a slab of constant properties and with its optical thickness as a free
parameter. \citet{truj05} showed that the observed intensity profiles and their ensuing $\cal R$ values can be reproduced by choosing an optical thickness significantly larger than unity. \citet{Centeno06} demonstrated that this optical thickness is related to the coronal irradiance (through the ratio $\cal R$), thus providing a physical meaning to the free parameter in the slab model {  (see also Centeno et al. 2007)}.

We present novel observations showing the spectral emission of \ion{He}{i} 10830~\AA\ and its dependence on the height of the
spicules above a quiet region. We compare the deduced observational $\cal R$ with that obtained from detailed non-LTE numerical calculations using available atmospheric profiles. 
%__________________________________________________________________

\section{Observations}
The observations were carried out on December 4$^{th}$, 2005, at the Vacuum
Tower Telescope (VTT) at Observatorio del Teide. They were supported by the
Kiepenheuer Adaptive Optics System \citep[KAOS,][]{luehe03}. {  We used the Echelle spectrograph of the VTT and the new version (TIP-2) of the Tenerife Infrared Polarimeter \citep{pilllet99}, which has a larger CCD camera \citep{collados07} }. The seeing conditions were good {  (average seeing after KAOS correction around 7 cm, maximum 12 cm)}.

{  The strong darkening close to the solar limb and the actual presence of the
limb make it difficult to use KAOS for off-limb observations, since the
correlation algorithm was not developed for this kind of observation.} Fortunately, we could use a nearby facula inside the limb for the tracking lock point of the adaptive optics system. 

Although we carried out full Stokes vector observations, in 
this study we have used only the intensity profiles. The spectral 
region covered by TIP-2 spans from 10826 to 10837~\AA\ and contains 
the \ion{He}{i} 10830~\AA\,  multiplet. {The spectral
sampling was  $11$ m\AA/px}, and the slit was 40\arcsec\, long and  0\farcs5\, wide. It was oriented parallel to the NE limb, in a region far from major activity at that time, at 59\fdg8 from North. We scanned the full height of the spicules, up to
7\arcsec\ off the visible solar limb,  with a step size of 0\farcs35.  

At each position, 5 consecutive spectra were measured with 10 accumulations of
250~ms each, with a total integration time of 2.5~s per off-limb slit
position. A nearby disc profile was also taken to help remove the scattered light from the off-limb spectra {  via the data analysis}. 

%%%%%%%%%%%%%%%%%%%%%%%%%%%%%%%%%%%%%%%%%%%%%%%%%%%%%%%%%%%%

\section{Data reduction and processing \label{ajuste}}

We applied usual flatfielding and dark current corrections to the data. A high frequency, quasi-periodic, spurious electronic pattern in the profiles
was removed using a low-pass Fourier filter, which left the frequencies containing the spectral line information untouched.

We define the position of the solar limb as the height of the first scanning position where the helium line appears in emission. For increasing distances to the solar limb a decreasing amount of sunlight is added by scattering in the Earth's atmosphere and by the telescope optical surfaces to the signal. Since the true off-limb continuum must be close to zero, i.e. below our detection limit, the observed continuum signal measures the spurious
light.  Therefore, we removed the spurious continuum intensity level by using the information given by a nearby average disc spectrogram. This first subtraction estimates the continuum level on a region 6 \AA \, away from the  \ion{He}{i} 10830~\AA \, emission lines. 
After this correction with a coarse estimate of the spurious light, a 
second correction was applied to remove the residual continuum level 
seen around the emission lines. This was needed since the transmission 
curve of the used prefilter is not flat but variable with wavelength.

\begin{figure}[h]
\center \includegraphics[width=0.5\textwidth]{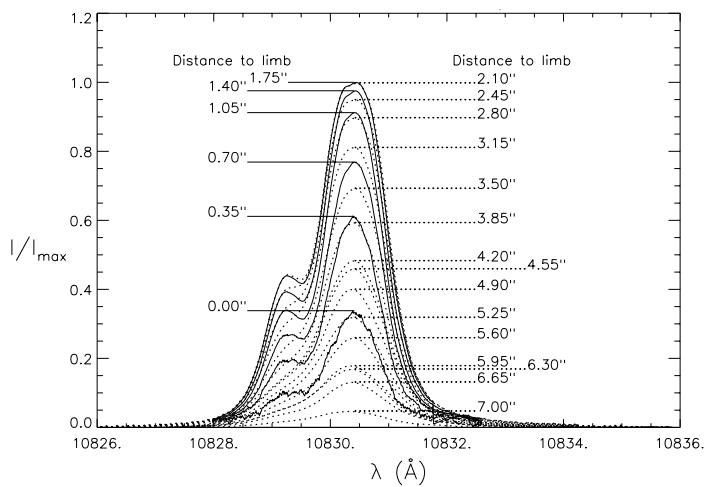} 
\caption{Measured \ion{He}{i} 10830~\AA\ emission profiles for increasing
  distances to the solar limb, scanning a broad range of the height extension of the 
  spicules. Each profile is the average of the 312 pixels along the slit (which was always kept parallel to the limb)}  
\label{fig:spe}
\end{figure}

\begin{figure}[h]
\center \includegraphics[width=0.47\textwidth]{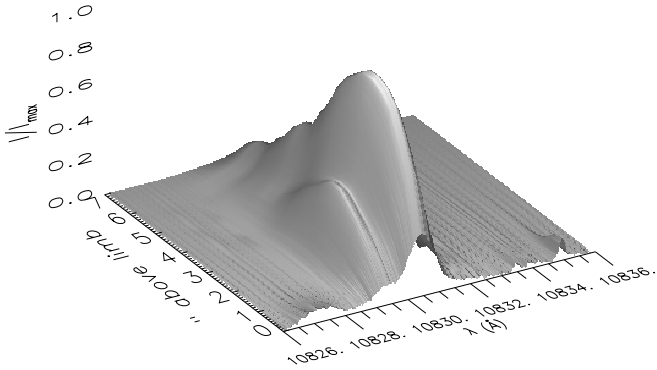} 
\caption{3D representation of the measured \ion{He}{i} 10830~\AA\ emission
  profiles for increasing distances to the solar visible limb. Note that the  x-axis is
  wavelength, the y-axis the height above solar limb and the z-axis the intensity
  normalised to the maximum emission in the line centre of the red component.}
\label{fig:3d}
\end{figure}

{  Figure~\ref{fig:spe} shows the emission profiles of the \ion{He}{i} 10830~\AA\, (after the reduction process) for different heights above the limb. 
Figure~\ref{fig:3d} illustrates this in three dimensions, as a function of
wavelength and the distance to the solar limb, clearly showing 
a change in the intensity ratio of the blue and red components of the 
multiplet $({\cal R}=I_{\rm blue}/I_{\rm red})$ with height. 
This will be discussed in Sect. \ref{results}.}

\begin{figure}[h]
\hspace{-0.3cm} \includegraphics[width=0.5\textwidth]{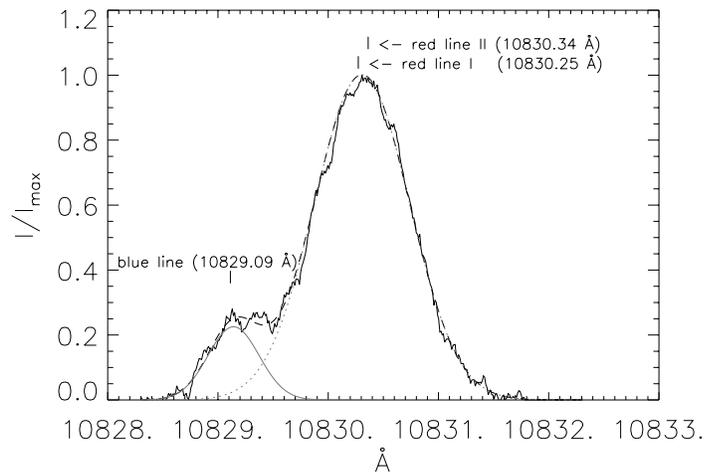} 
\caption{Determination of the blue and red components of the \ion{He}{i} 
10830~\AA\ triplet from the observed emission profiles. In this example the
 slit was placed at 1\farcs4 off the solar visible limb. See text for details. 
  The solid line {  represents} the average emision profile. The dotted line is the Gaussian fit to the symmetrised red
  component. Subtraction of this from the observed profile leaves the blue
  component, which is also fitted by a Gaussian profile (thin solid line). 
The sum of both Gaussians (dashed line)  gives the fit to the 
observed profile.}
\label{fig:ajuste}
\end{figure}

For the calculation of $\cal{R}$ we need to determine the amplitudes of the 
blue and red components of the emission profile (as shown in Fig.  
\ref{fig:ajuste}).

To determine the core wavelength of the red 
component of the triplet we fitted a Gaussian profile to its core, in a 1.3 \AA\ 
range around the maximum. After symmetrising the observed profile around this 
maximum, using the values on the red side of the red component, we fitted another Gaussian 
function to the resulting symmetric profile. Subtraction of the fitted 
symmetric profile from the data leaves the emission profile of the blue 
component, which was also approximated by a Gaussian to determine its central wavelength.  Our tests trying to fit directly both profiles using two Gaussians failed in a number of cases, probably due to the following reasons: (a) the red component is in fact the result of two blended lines, (b) the much weaker blue component was almost hidden in the broadened red component, and (c) the presence of noise. Our technique determines first the red component and then, after subtraction of the fitted profile, the blue one.

We have thus separated the helium emissions into their red and blue
components  assuming only 2 that both are present and that they are both symmetric. We can now measure their widths and intensities and also check that the line core positions coincide with the theoretical ones. After the fitting process the residuals
between measured and observed profiles were small, the largest errors occurring
in the determination of the core intensities of the red line. This happens because the red component consists of two blended lines (with a separation of 0.09 \AA), a fact that flattens the emission profile near the core as opposed to a more peaked Gaussian function. Nevertheless, the differences between fitted profiles
and data are only significant in the red core and are always lower than
+0.08 of the maximum normalised intensity, with a mean difference of $\sim$0.02. To avoid systematic errors, we used the observational values for the centre of
the red component when calculating $\cal R$.

%%%%%%%%%%%%%%%%%%%%%%%%%%%%%%%%%%%%%%%%%%%%%%%%%%%%%%%%%%%%%%%%%%%%%%%%%

\section{Results\label{results}}

The chromospheric temperature and density are too low  to populate the
ortho-helium levels via collisions \citep{avrett94}. The EUV irradiation from
the corona (CI) ionises the para-helium, and  
the subsequent recombinations lead to an overpopulation of all the 
ortho-helium levels, in particular those involved in the 10830~\AA\
transitions. \citet{Centeno06} and Centeno et al. (2007) have modelled the off-the-limb emission profiles and concluded that the ratio $\cal{R}$\,=\,$I_{blue}/I_{red}$ is a function of the height and a direct
tracer of the amount of CI. Here we compare the results from the theoretical modelling with observations.

\begin{figure}[h]
\center \includegraphics[width=0.48\textwidth]{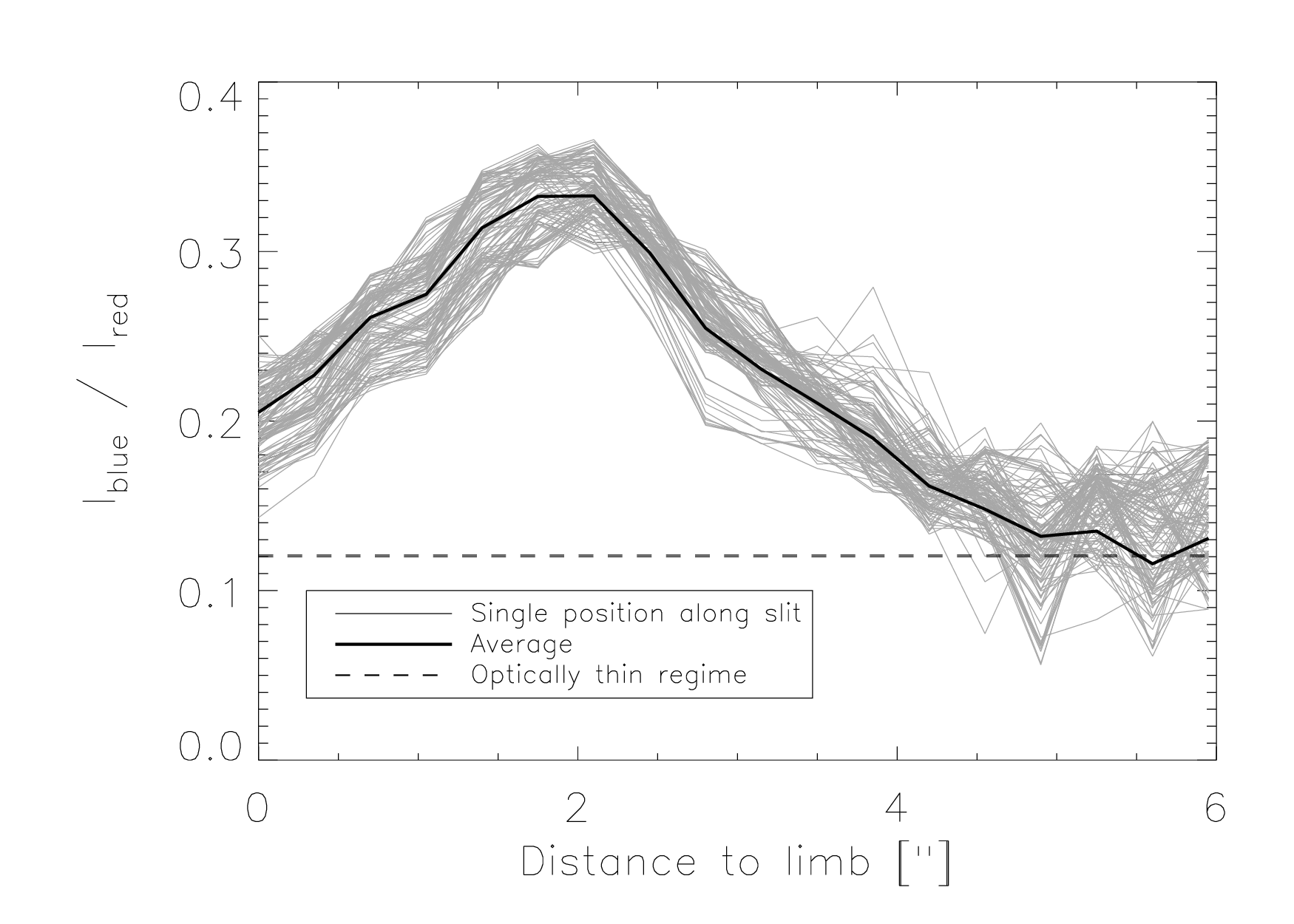}
%\vspace{-0.5cm} 
\caption{Measured ratio $\cal R$\,=\,$I_{blue}/I_{red}$ as a function of
  distance to the solar limb. Thin lines come from
  each pixel along the slit. The thick solid line represents the average and the dashed line the value of the optically thin regime.} 
\label{fig:ratios}
\end{figure}

\citet{truj05} {modelled their spectropolarimetric observations assuming a slab with constant physical properties with a given optical thickness}. In the optically thin regime $\cal R$\,$\sim$\,0.12, which is the ratio of the relative oscillator strengths of the triplet. As the optical
thickness (at the line-centre of the red blended component) grows, this ratio also increases 
until it reaches a saturation value slightly larger than 1 for $\tau\sim10$. (This type of calculation can be done and improved as explained in Trujillo Bueno \& Asensio Ramos 2007). To reproduce the observed emission profile \citet{truj05} had to choose $\tau \sim 3$. Interestingly, {  the values of $\tau$ yielded by this modelling strategy are consistent} with the more realistic 
approach of Centeno (2006), where non-LTE radiative transfer calculations in 
semi-empirical models of the solar atmosphere are presented, using spherical 
geometry and taking into account the ionising coronal irradiation.
With our data we are able to test such theoretical calculations by comparing 
the measured values of $\cal R$ with those resulting from various chromospheric
models. This way we may eventually trace the amount of CI inciding on the 
spicules. The analysis described in Sect.~\ref{ajuste} yielded the 
values of $\cal R$ for the observed profiles. The resulting dependence on 
the distance to the solar
limb, for each pixel along the slit and each position of the slit above the
limb, are presented in Fig.~\ref{fig:ratios}. The solid black line gives the average value of $\cal R$.

The dependence of $\cal R$ with height can be understood in a qualitative way as
follows: in the outer layers of the chromosphere the density is so low 
that the transitions occur in the optically thin regime. 
With decreasing altitude the ratio $\cal R$ increases (proportionally with 
density) until a maximum optical thickness is reached.  At even lower 
layers, although the density still continues to rise,  the extinction of the coronal
irradiance leads to a reduction in the number of ionizations, which results 
in a decrease of the optical thickness in the core wavelength of the red 
component, {  and thus in a decrease of $\cal R$.}

For a quantitative comparison with theoretical modelling we have 
used the results from \citet{Centeno06} and \cite{cente07} where they calculated the ratios $\cal R$ for different
standard model atmospheres: FAL-C and FAL-P \citep{fontenla91} and 
FAL-X \citep{avrett95}. The FAL-C and FAL-X models may be considered as illustrative of the thermal conditions in the quiet Sun, while the FAL-P model of a plage region. The FAL-X model has a relatively cool atmosphere in order to explain the molecular CO absorption at 4.6~$\mu$.

\begin{figure}[h]
\hspace{-0.5cm}\includegraphics[width=0.5\textwidth]{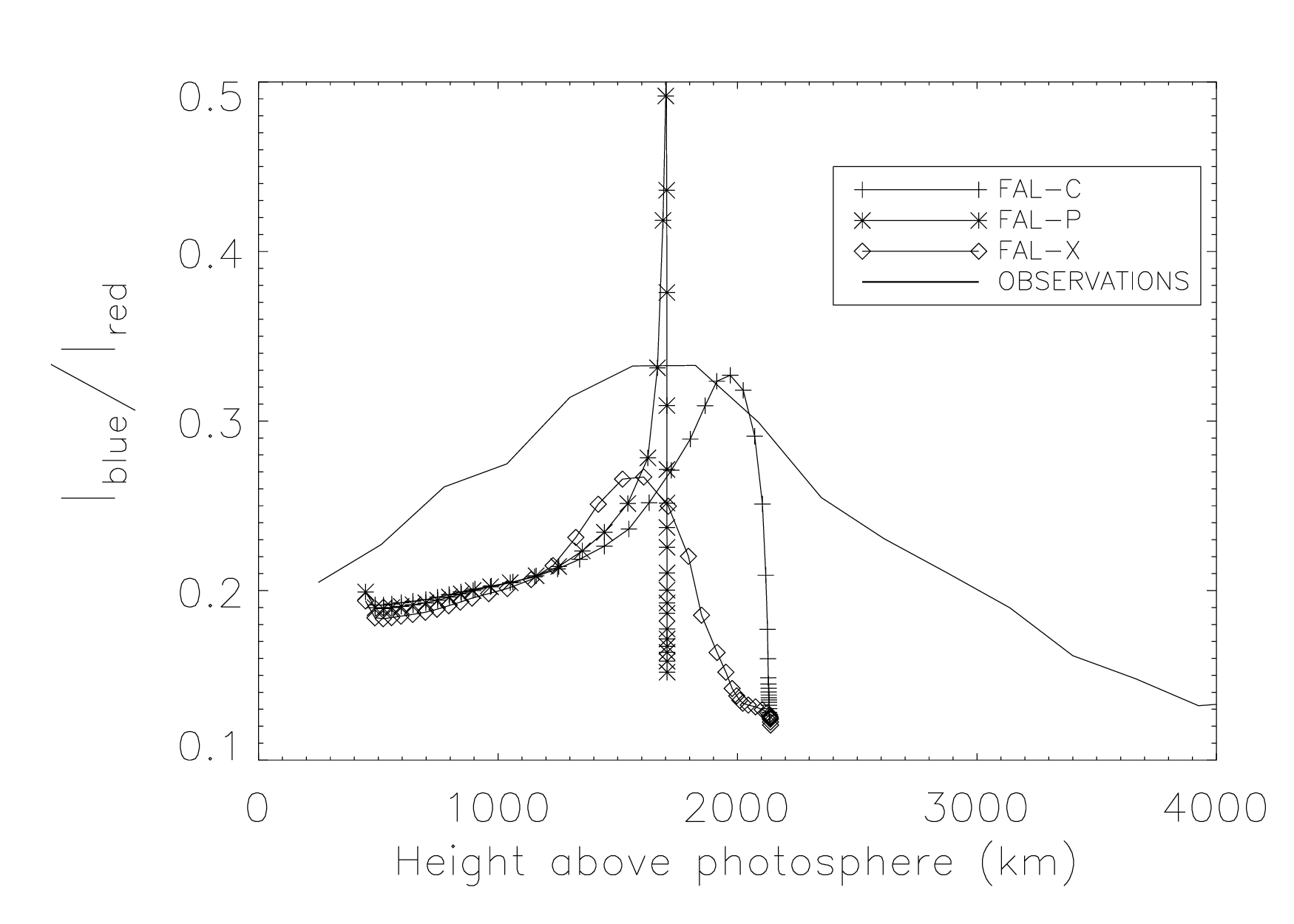} 
\caption{Observed (average) vs. theoretical variation of the ratio $\cal R$$ =
  I_{blue}/I_{red}$ 
  with height.}
\label{fig:comp}
\end{figure}

The comparison is shown in Fig.~\ref{fig:comp}. We notice that the modelled
height variations of $\cal{R}$ agree only in a qualitative manner with what is 
found in our observations. However, the calculations from different models of 
the solar atmosphere are unable to reproduce the measured ratio. Higher values of the coronal irradiance lead to an increase of the optical thickness (at the line centres of the \ion{He}{i} multiplet) and an upward shift in the run of $\cal{R}$ vs. height. Yet the shape of the height dependence is mainly given by the atmospheric density profile and the attenuation of the ionising radiation as it reaches the lower layers of the chromosphere.
It is also clear from Fig.~\ref{fig:comp} that the models do not extend high enough. 

%%%%%%%%%%%%%%%%%%%%%%%%%%%%%%%%%%%%%%%%%%%%%%%%%%%%%%%%%%%%%%%%%%%%%

\section{Conclusions\label{conclusion}}
The theoretical behaviour of the ratio $\cal R$ agrees
qualitatively with observations. Yet, a quantitative comparison shows poor
agreement. Also, the simulated ratios are highly model dependent. 
As already explained, {  the failure to reproduce the observed profiles is very likely due to the density stratification not being adequate for spicule modelling and to the limited vertical extension of the atmospheric models.}
Modelling of the intensity ratio $\cal R$ in the
\ion{He}{i} infrared triplet should account for the fact that the solar
chromosphere is inhomogeneous on small scales and that the spicules are
small-scale intrusions of chromospheric matter into the hot corona.

New data of spicule regions near the poles and the equator, below coronal 
holes or coronal active regions should help us to understand the detailed 
behaviour of the \ion{He}{i} 10830~\AA\ lines.
In further work, we will extend this study to the full Stokes vector, in order to see  
the variation of the linear polarization - or even the variation of the Hanle effect - with height. {  It would also be interesting to use the most recent models of active region fibrils and spicules \citep[e.g.,][]{hegg07} in order to see whether or not they agree with our observations.} Future models of the solar chromosphere should be constrained by the observational evidences presented here.

%%%%%%%%%%%%%%%%%%%%%%%%%%%%%%%%%%%%%%%%%%%%%%%%%%%%%%%%%%%%%%%%%%%%%

\begin{acknowledgements}
We thank M. Collados for the extensive help and discussions, as
well as for the reduction software. The help from A. Lagg was very
valuable during the reduction phase. BSAN acknowledges a PhD fellowship at the International Max Planck Research School {\em On Physical Processes in the Solar System and Beyond}. KGP thanks the 
Deutsche Forchungsgemeinschaft
for support through grant KN 152/29. RC and JTB 
acknowledge the support of the Spanish Ministerio de Educaci\'on y Ciencia through project AYA2004-05792. JTB thanks the Akademie der Wissenschaften zu G\"ottingen for the Gauss-Professur during a sabatical stay at the Institut f\"ur Astrophysik of the University 
of G\"ottingen. The Vacuum Tower Telescope is operated by the Kiepenheuer-Institut f\"ur Sonnenphysik, Freiburg, at the Spanish Observatorio del Teide of the Instituto de Astrof\'\i sica de Canarias. The National Center for Atmospheric Research is sponsored by the National Science Foundation
\end{acknowledgements}

%\bibliographystyle{aa}
%\bibliography{helio}
%\input{ref.tex}
%\bibliography{helio}

%\end{document}

\end{document}